\documentclass[aps,prb,showpacs,reprint,amsmath,amssymb,floatfix]{revtex4-1}
\usepackage{epsf}
\usepackage{graphicx}
\usepackage{amsmath}
\usepackage{amsfonts}
\usepackage{amssymb}
\usepackage{dcolumn}
\usepackage[version=3]{mhchem}
\usepackage{color,ulem}
\usepackage{bm}

\begin{document}

\title{Strain-engineered magnetic order in (LaMnO$_{3}$)$_n$/(SrMnO$_{3}$)$_{2n}$ superlattices }

\author{$^{1,2,3}$Qinfang Zhang} 
\email{qfangzhang@gmail.com}
\author{$^4$Shuai Dong}
\author{$^1$Baolin Wang} 
\author{$^{2,3,5}$Seiji Yunoki}

\affiliation{
$^1$Key Laboratory for Advanced Technology in Environmental Protection of Jiangsu Province, 
Yancheng Institute of technology,Yancheng 224051, China \\
$^2$Computational Condensed Matter Physics Laboratory, RIKEN ASI, Wako, Saitama 351-0198, Japan \\
$^3$CREST, Japan Science and Technology Agency, Kawaguchi, Saitama 332-0012, Japan \\
$^4$Department of Physics, Southeast University, Nanjing 211189, China \\
$^5$Computational Materials Science Research Team, RIKEN AICS, Kobe, Hyogo 650-0047, Japan
}

\date{\today}

\begin{abstract} 
Using first-principles calculations based on the density functional theory, we show a strong strain dependence of magnetic order in 
(LaMnO$_{3}$)$_n$/(SrMnO$_{3}$)$_{2n}$ (001) superlattices with $n=1,2$. 
The epitaxial strain lifts the degeneracy 
of Mn $e_{g}$ orbitals, thus inducing an inherent 
orbital order, which in turn strongly affects the ferromagnetic double exchange of itinerant $e_{g}$ electrons, competing with the 
antiferromagnetic superexchange of localized $t_{2g}$ electrons. For the case of tensile strain induced by SrTiO$_3$ 
(001) substrate, we find that the ground state is A-type antiferromagnetic and $d_{x^2-y^2}$ orbital ordered, which is in excellent agreement with recent 
experiments [S.~J.~May {\it et al.}, Nature Materials {\bf 8}, 892 (2009)]. 
Instead, for the case of compressive strain induced by LaAlO$_3$ (001) substrate, we predict that the ground state is C-type antiferromagnetic and 
$d_{3z^2-r^2}$ orbital ordered.

\end{abstract}


\maketitle

\section{Introduction}
Transition metal oxides in perovskite based structures exhibit a wide variety of phases with different electronic, magnetic, 
and orbital structures, and show rich functionalities such as high-$T_{\rm C}$ superconductivity, 
colossal magnetoresistance, and multiferroics.~\cite{imada@rmp98} A recent advance in epitaxial growth 
techniques has made it even possible to fabricate transition metal oxide heterostructures with sharp and smooth interfaces 
controlled at the atomic scale.~\cite{ahn@rmp06} 
In these heterostructures, many unique properties, not found in the corresponding alloy compounds made of the same 
composite elements, have been observed, which include e.g., two dimensional electron gas with high 
mobility at the heterostructure interfaces,~\cite{ohtomo@nature06} indicating the promising potential of oxide heterostructures 
for future technological applications.~\cite{hwang@nm12}

In the case of manganites,~\cite{dagotto@pr01} LaMnO$_{3}$ is an A-type antiferromagnetic insulator and 
SrMnO$_3$ is a G-type 
antiferromagnetic insulator. On one hand, the randomly-cation-doped alloy La$_{1-x}$Sr$_x$MnO$_3$ exhibits a rich 
magnetic phase 
diagram, depending on the doping concentration $x$. On the other hand, La/Sr cation-ordered analogues forming 
superlattices behave quite differently from their alloy 
compounds.~\cite{adamo@apl08,zhao@prl08,bhattacharya@prl08,perucchi@nl10,galdi@prb12} 
For example, La$_{2/3}$Sr$_{1/3}$MnO$_3$ alloy has a mixed valence of Mn$^{3+}$/Mn$^{4+}$, and the ground state is 
ferromagnetic half metallic due to the double exchange mechanism.~\cite{dagotto@pr01} 
To the contrary, it is found experimentally that cation-ordered 
(LaMnO$_3$)$_{2n}$/(SrMnO$_3$)$_n$ (001) superlattices are insulating when $n$ is larger 
than 3.~\cite{adamo@apl08,bhattacharya@prl08} 
This change of behavior 
is easily understood because the number $n$ of SrMnO$_3$ layers control the quantum confinement potential: when $n$ is 
small, the confinement potential is small and the $e_g$ electrons are distributed uniformly, thus expecting the phases 
similar to the alloy La$_{1-x}$Sr$_x$MnO$_3$.  When $n$ is large, the confinement potential becomes large enough 
to trap the $e_g$ electrons in LaMnO$_3$ layers, and thus the bulk properties of LaMnO$_{3}$ and SrMnO$_3$ would 
be observed. Several theoretical studies for (LaMnO$_3$)$_{2n}$/(SrMnO$_3$)$_n$ superlattices have been reported 
to understand their electronic and magnetic properties.\cite{dong@prb08,nanda@prb09}

More recently, Bhattacharya {\it et al.}~\cite{bhattacharya@arl07,may@nm09} have experimentally studied the transport and the magnetic properties of similar 
superlattices (LaMnO$_3$)$_n$/(SrMnO$_3$)$_{2n}$ grown on SrTiO$_3$ (001) substrate. They have found that the ground state 
of these superlattices with $n=1,2$ are A-type antiferromagnetic metals with N\'{e}el temperature ($T_{\rm N}$) which is higher than 
that observed in any alloy La$_{1-x}$Sr$_x$MnO$_3$ compound.~\cite{may@nm09} Although the similar physical principles 
found in (LaMnO$_3$)$_{2n}$/(SrMnO$_3$)$_n$ superlattices are certainly expected to apply here, the systematic theoretical 
investigations are required to understand the main ingredients which determine the electronic as well as the magnetic properties of 
(LaMnO$_3$)$_n$/(SrMnO$_3$)$_{2n}$ superlattices. 

Here, in this paper, performing first-principles calculations based on the density functional theory, we study the electronic 
and the magnetic structures of (LaMnO$_3$)$_n$/(SrMnO$_3$)$_{2n}$ (001) superlattices with $n=1,2$. We show 
that the magnetic properties are governed not only by the quantum confinement potential caused by periodic alignment 
of cation ions La$^{3+}$/Sr$^{2+}$ 
but also by the strain induced by substrates on which 
the superlattices are grown. Namely, for the case of tensile strain induced by SrTiO$_3$ (STO) (001) substrate, our calculations 
show that the ground state of these superlattices are A-type antiferromagnetic and $d_{x^2-y^2}$ orbital ordered with 
higher $T_{\rm N}$ for $n=1$ than for $n=2$. This is indeed in excellent agreement with recent experimental 
observations.~\cite{may@nm09} 
Instead, for the case of compressive strain induced by LaAlO$_3$ (LAO) (001) substrate, we predict C-type antiferromagnetic 
and $d_{3z^2-r^2}$ orbital orders with higher $T_{\rm N}$ for $n=1$ than for $n=2$.

The rest of this paper is organized as follows. After describing the computational details in Sec.~\ref{method}, the numerical 
results for the cases of SrTiO$_3$ substrate and LaAlO$_3$ substrate are presented in Sec.~\ref{results_sto} and Sec.~\ref{results_lao}, 
respectively, followed by discussion of the confinement potential in Sec.~\ref{confine}. 
Sec.~\ref{summary} summaries this paper.

\section{Computational Methods \label{method}}

We perform the first-principles electronic structure calculations based on the projected augmented wave pseudopotentials 
using the 
Vienna Ab initio Simulation Package (\textit{VASP}).~\cite{kresse@prb93,kresse@prb96} The valence states include 3p4s3d 
and 2s2p for Mn and O, respectively. The electron interactions are described using the generalized gradient 
approximation (GGA) and  the rotationally invariant GGA+\textit{U} method~\cite{blochl@prb94,kresse@prb99,dudarev@prb98} with the effective $U_{\rm eff}$, i.e., $U-J$, from 1 eV to 5 eV for \textit{d} electron states.  
Compared to the GGA,  the GGA+$U$ approach gives an 
improved description of \textit{d} electron localization.\cite{anisimov@jpcm97} 
The atomic positions of superlattices are fully optimized iteratively until the Hellman-Feynman forces 
are 0.01 eV/$\text{\AA{}}$ or less. 
The plane-wave cutoff is set to be 500 eV and a $12\times12\times12$  Monkhorst-Pack 
k-point grid is used in combination with the tetrahedron method.~\cite{blochl@prb94a} 

\begin{figure*}[ht]
\begin{center}
\includegraphics[width=36pc]{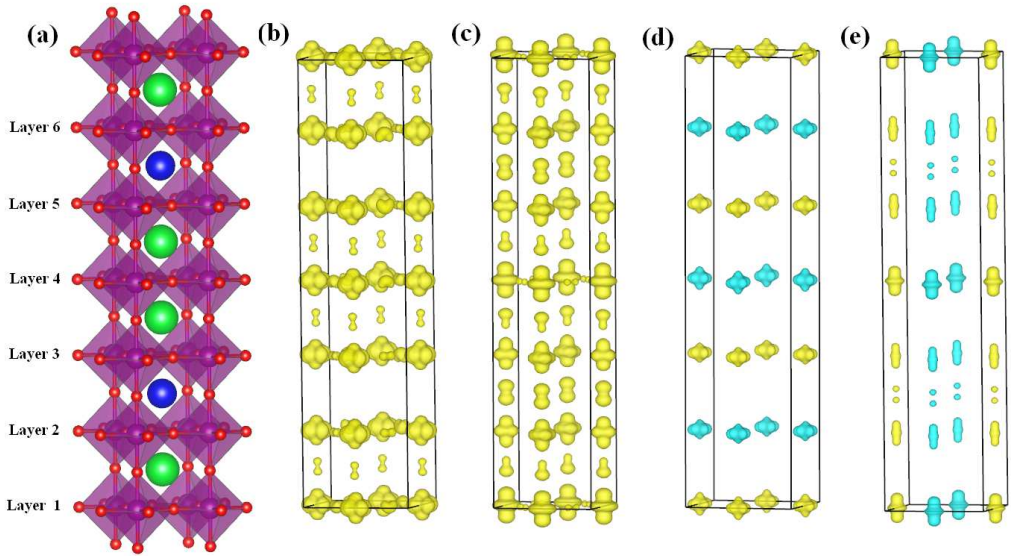}
\end{center}
\caption{(Color online). (a) A schematic figure of the supercell considered for LaMnO$_{3}$/(SrMnO$_{3}$)$_2$ (001) 
superlattices, and the projected charge [(b) and (c)] and spin [(d) and (e)] density distributions 
(integrated from Fermi level down to $-0.5$ eV using GGA) for LaMnO$_{3}$/(SrMnO$_{3}$)$_2$ superlattices grown on 
SrTiO$_3$ [(b) and (d)]  and LaAlO$_3$ [(c) and (e)] (001) substrates. 
The loci of MnO layers are indicated in (a), where red, blue, green, and purple spheres indicate O, La, 
Sr, and Mn atoms, respectively. In (d) and (e), the up and down spin densities are denoted by yellow and light blue, respectively.
}
\label{orbital}
\end{figure*}

\begin{figure*}[ht]
\begin{center}
\includegraphics[width=36pc]{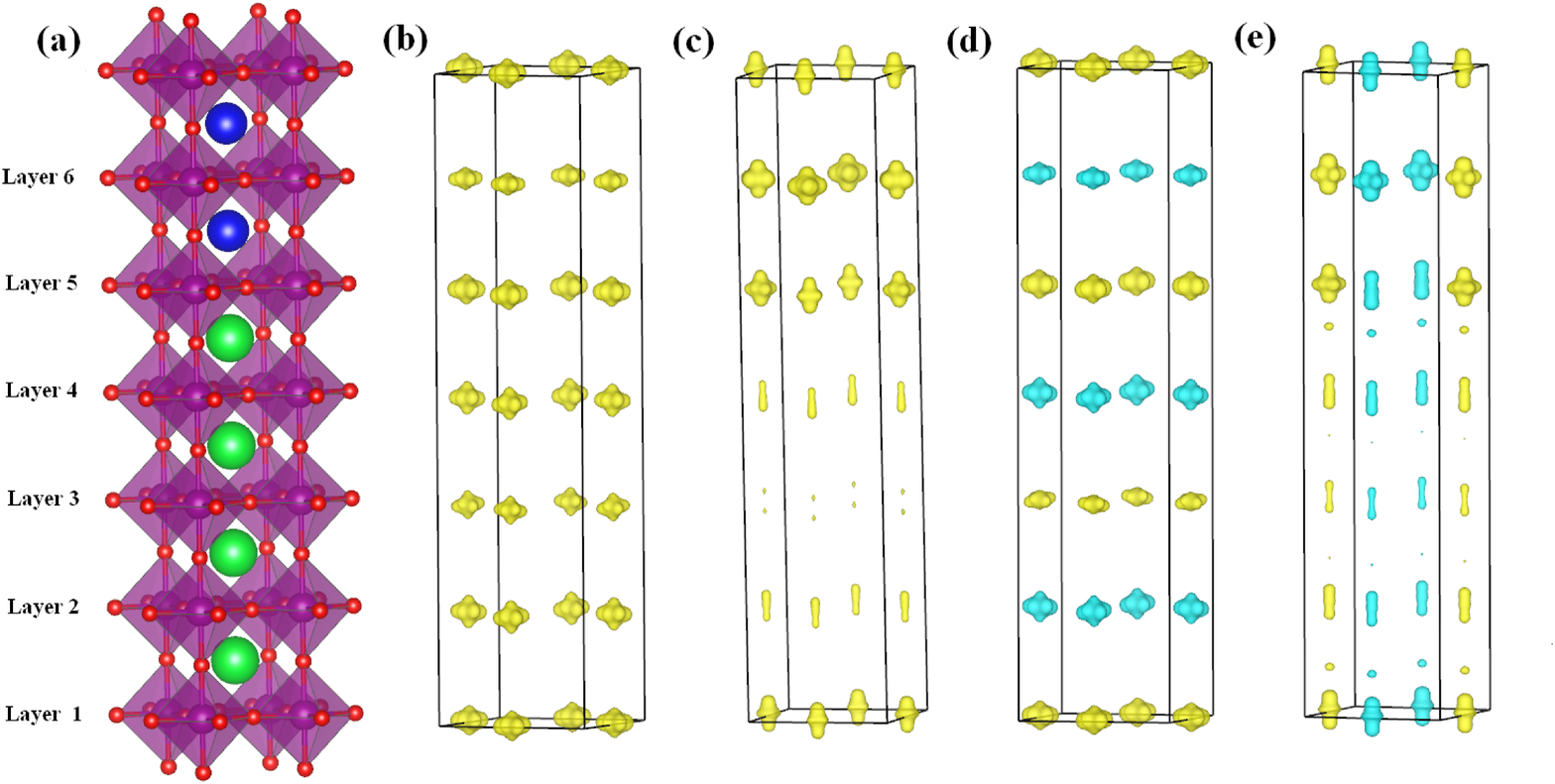}
\end{center}
\caption{(Color online). 
(a) A schematic figure of the supercell considered for (LaMnO$_{3}$)$_2$/(SrMnO$_{3}$)$_4$ (001) 
superlattices, and the projected charge [(b) and (c)] and spin [(d) and (e)] density distributions
(integrated from Fermi level down to $-0.5$ eV using GGA) for (LaMnO$_{3}$)$_2$/(SrMnO$_{3}$)$_4$ superlattices 
grown on SrTiO$_3$ [(b) and (d)]  and LaAlO$_3$ [(c) and (e)] (001) substrates. 
The loci of MnO layers are indicated in (a), where red, blue, green, and purple spheres indicate O, La, 
Sr, and Mn atoms, respectively. In (d) and (e), the up and down spin densities are denoted by yellow and light blue, respectively.
}
\label{orbital2}
\end{figure*}

\begin{figure*}[ht]
\begin{center}
\includegraphics[width=30pc]{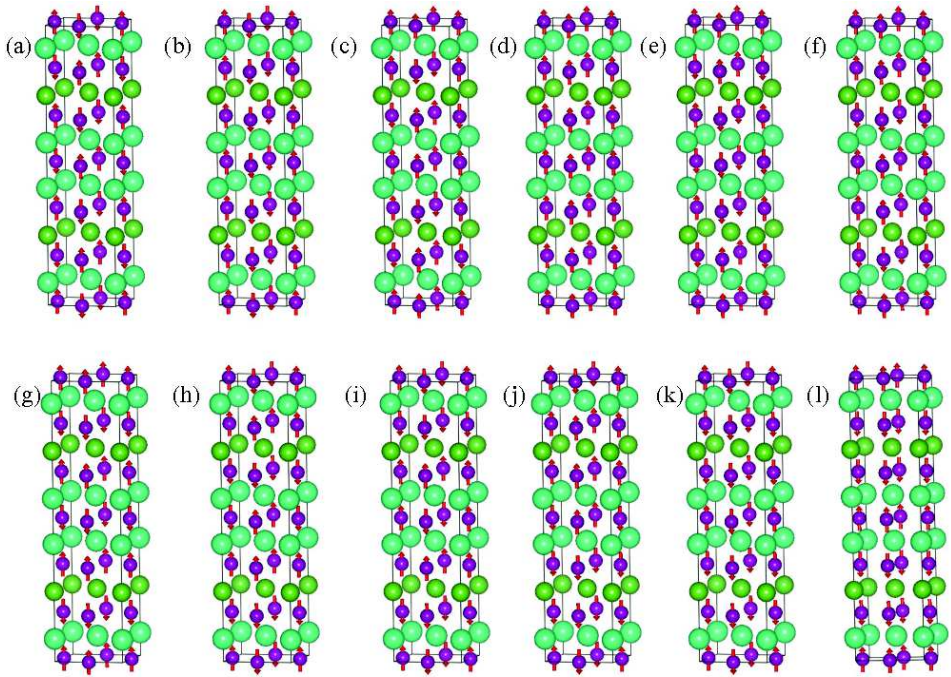}
\end{center}
\caption{\label{crystal-La1} (Color online). 
12 different magnetic structures considered for LaMnO$_{3}$/(SrMnO$_{3}$)$_2$ superlattices: G-AFM (a), C-AFM (b), M1-AFM (c),
 FM (d), M2-AFM (e), D-AFM (f), A-AFM (g), M3-AFM (h), M4-AFM (i), M5-AFM (j), M6-AFM (k),  and D1-AFM (l). Mn spins are 
 indicated by arrows. Aqua, lime, and violet spheres stand for Sr, La, and Mn atoms, respectively. O atoms are omitted for clarity.}
\end{figure*}

\begin{figure*}[ht]
\begin{center}
\includegraphics[width=30pc]{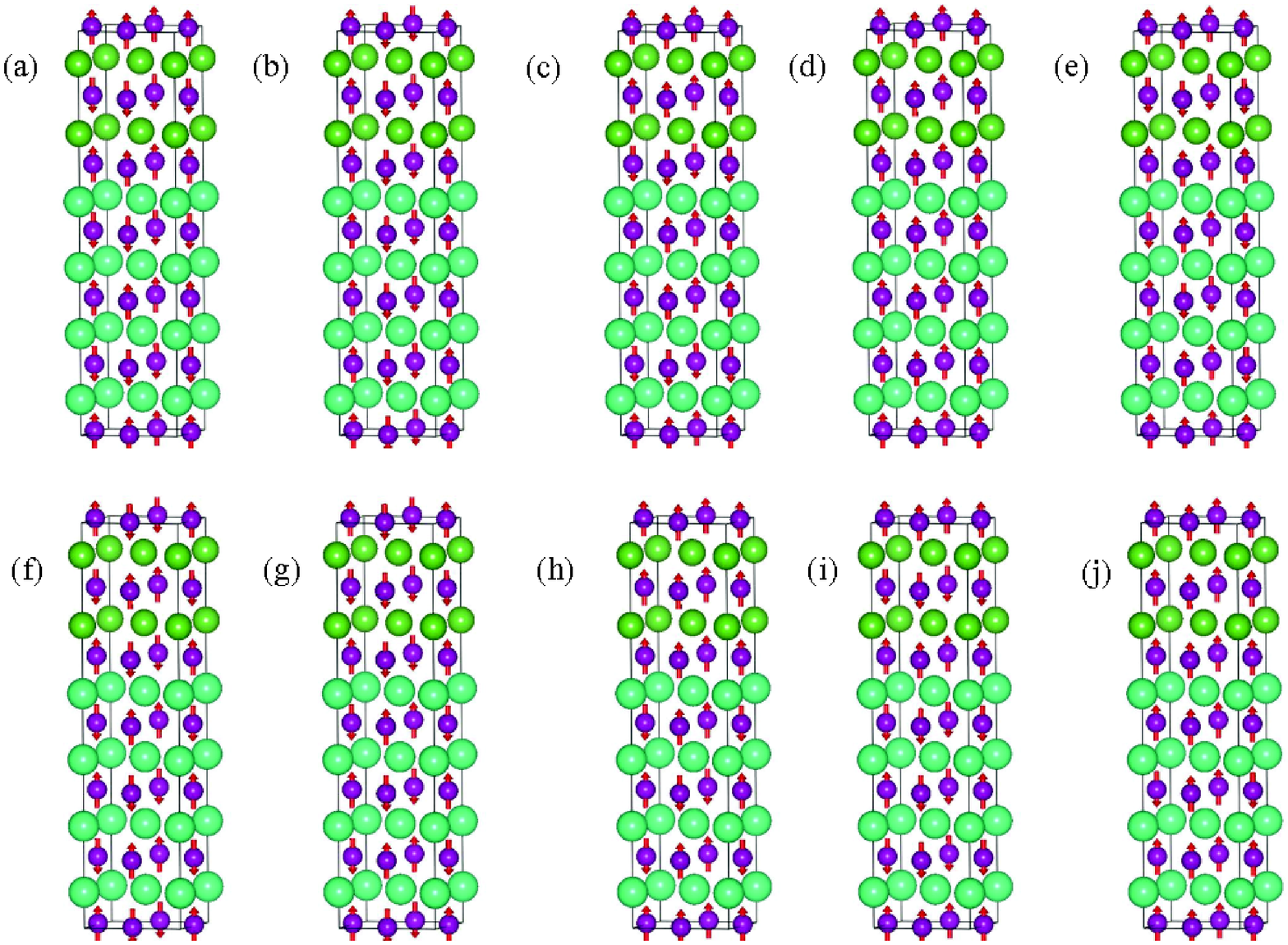}
\end{center}
\caption{\label{crystal-La2} (Color online). 
10 different magnetic structures considered for (LaMnO$_{3}$)$_2$/(SrMnO$_{3}$)$_4$ superlattices: A-AFM (a), C-AFM (b), 
D-AFM (c), FM (d), M2-AFM (e), G-AFM (f), M1-AFM (g), M3-AFM (h), M4-AFM (i), and M5-AFM (j). 
Mn spins are indicated by arrows. Aqua, lime, and violet spheres stand for Sr, La, and Mn atoms, respectively. O atoms are omitted 
for clarity.
}
\end{figure*}

The supercells considered here consist of 6 MnO$_2$ layers, 2 LaO layers, and 4 SrO layers for both $n=1$ and 2, 
as shown in Fig.~\ref{orbital} (a) and Fig.~\ref{orbital2} (a).  
We consider 12 and 10 different magnetic moment alignments to search for the ground state magnetic structures for 
LaMnO$_{3}$/(SrMnO$_{3}$)$_2$ and (LaMnO$_{3}$)$_2$/(SrMnO$_{3}$)$_4$ superlattices, as shown in 
Fig.~\ref{crystal-La1} and Fig.~\ref{crystal-La2}, respectively. 
These magnetic structures include not only simple ferromagnetic, A-type, C-type, and G-type antiferromagnetic 
structures~\cite{wollan} but also magnetic structures with mixed combinations of these simple magnetic structures. 
The epitaxial constraint on these superlattices, which is grown on substrates, is to fix the in-plane lattice constants. Thus, to simulate the 
strain effect, we fix the in-plain lattice constants ($a$) of the superlattices to the ones of 
substrates, i.e., $a=3.905$ \AA{} for SrTiO$_3$ substrate~\cite{mitchell} and $a=3.81$ \AA{} for LaAlO$_3$ substrate, \cite{nakatsuka} 
and the lattice constant ($c$) perpendicular to MnO$_2$ layers is fully relaxed. Atomic positions are also fully optimized.

\section{Results}

\subsection{\label{results_sto}(LaMnO$_3$)$_n$/(SrMnO$_3$)$_{2n}$ on SrTiO$_3$}

\begin{figure}[ht]
\begin{center}
\includegraphics[width=10pc]{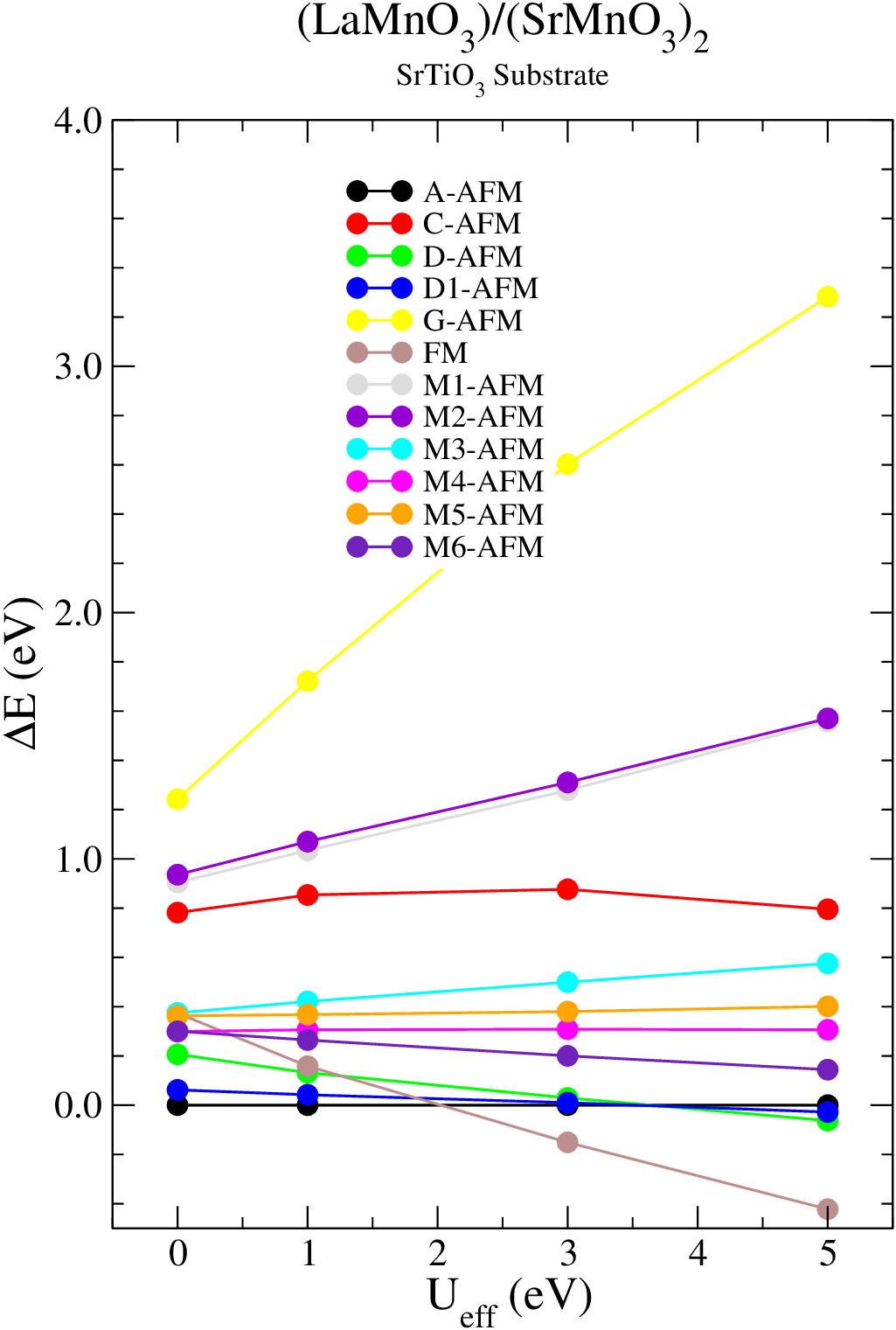}
\includegraphics[width=10pc]{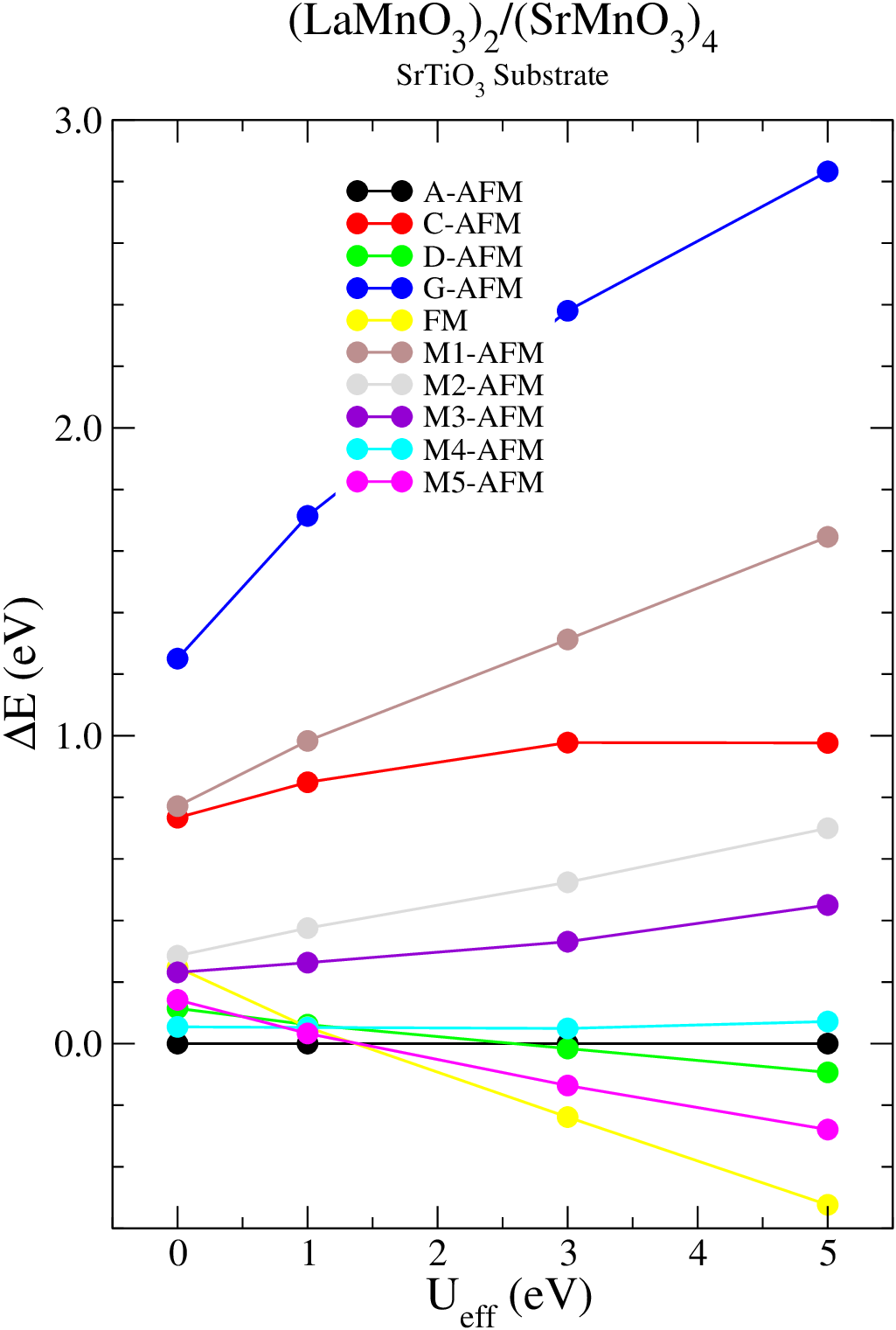}
\end{center}
\caption{\label{La-Ti} (Color online). 
$U_{\rm eff}$ dependence of the relative energies (calculated using GGA+$U$) for various magnetic structures 
(see Fig.~\ref{crystal-La1} and Fig.~\ref{crystal-La2}) compared to A-type 
antiferromagnetic state for (LaMnO$_{3}$)$_n$/(SrMnO$_{3}$)$_{2n}$ with $n=1$ (left) and 
$n=2$ (right) on SrTiO$_3$ substrate. 
}
\end{figure}

Let us first examine (LaMnO$_3$)$_n$/(SrMnO$_3$)$_{2n}$ (001) superlattices on SrTiO$_3$ (001) substrate. 
Our systematic GGA calculations reveal that the ground states of these superlattices with $n=1$ and 2 are both A-type 
antiferromagnetic metals. A schematic spin alignment of A-type antiferromagnetic order is shown in Fig.~\ref{crystal-La1} (g) 
and Fig.~\ref{crystal-La2} (a). Indeed, as shown in Fig.~\ref{orbital} (d) and 
Fig.~\ref{orbital2} (d), the projected spin density distribution, calculated by integrating spin density of 
occupied states from Fermi level down to $-0.5$ eV, clearly indicates the A-type antiferromagnetic spin order. 
Our GGA+$U$ calculations also find that these A-type antiferromagnetic states are robust against electron correlations, 
and they are indeed stable up to $U_{\rm eff}=2$ eV 
for $n=1$ and $U_{\rm eff}=1.3$ eV for $n=2$ (see Fig.~\ref{La-Ti}).~\cite{note}

Since the supercell sizes and the numbers of each type of atoms are the same, we can simply compare the total energy of 
these two different superlattices. Tab.~\ref{table_energy} summarizes the total energies for the A-type antiferromagnetic 
states and other magnetic states. Since the A-type (C-type) magnetic structure is ferromagnetic (antiferromagnetic) 
within the $ab$ plane and antiferromagnetic (ferromagnetic) along the $c$ direction, we can approximately estimate an effective 
magnetic exchange ($J_{\rm eff}$) simply by comparing the total energy of the A-type and the C-type antiferromagnetic states. 
It is clearly observed in Tab.~\ref{table_energy} that the stabilization energy of the A-type 
antiferromagnetic state, i.e., $J_{\rm eff}$, is larger for $n=1$ than for $n=2$. This implies that $T_N$ for $n=1$ is higher than that for 
$n=2$. These results are in excellent agreement with experimental observations by May {\it et al.}~\cite{may@nm09}

\begin{table}[htbp]
\caption{\label{table_energy} Total energies (in unit of eV) of (LaMnO$_{3}$)$_n$/(SrMnO$_3$)$_{2n}$ 
superlattices ($n=1,2$) calculated using GGA.  
FM, A-AFM, and C-AFM stand for ferromagnatic, A-type antiferromagnetic, and C-type antiferromagnetic states, respectively. 
 }

\begin{tabular}{|c|c|c|c|c|c|c|}
\hline
  \textit{n} & \multicolumn{3}{|c|}{SrTiO$_3$ substrate}  & \multicolumn{3}{|c|}{LaAlO$_3$ substrate}\\\cline{2-7}
  &FM&A-AFM&C-AFM&FM&A-AFM&C-AFM\\  \hline
 \textit{n}=1&  $-466.262$   & $-466.636$  & $-465.854$  & $-465.852$   & $-465.459$ & $-466.466$  \\
\hline
\textit{n}=2&  $-465.637$   & $-465.885$  & $-465.151$ & $-465.398$  & $-465.015$  &  $-465.858$ \\
\hline
\end{tabular}
\end{table}

Since the epitaxial constraint of substrates is to fix the in-plane lattice constant $a$ of the superlattices, the tetragonal distortion 
should inevitably occur, which in turn affects the relative occupation of Mn $e_g$ electrons. 
Indeed, as shown in Tab.~\ref{table_lattice}, we find that the SrTiO$_3$ substrate induces tensile strain with $a >c$, in 
which $d_{x^2-y^2}$ orbital is lower in energy than $d_{3z^2-r^2}$ orbital. This can be seen in the projected charge density 
distribution, the integrated charge density from Fermi level down to $-0.5$ eV, shown in Fig.~\ref{orbital} (b) and 
Fig.~\ref{orbital2} (b), indicating that $e_g$ electrons preferably occupy $d_{x^2-y^2}$ orbital. 
Because of this orbital order induced inherently by the substrate strain, the A-type antiferromagnetic order is stabilized. 
Remember that the magnetic interaction between Mn ions is determined by competition between the ferromagnetic 
double exchange via itinerant Mn $e_g$ electrons and the antiferromagnetic superexchange between localized Mn 
$t_{2g}$ electrons. When $d_{x^2-y^2}$ orbital is occupied rather than $d_{3z^2-r^2}$ orbital, the strong double exchange 
induces ferromagnetic order in the $ab$ plane while the weak itineracy of $d_{x^2-y^2}$ electrons along the $c$ direction 
reduces substantially the double exchange and as a results the superexchange between $t_{2g}$ electrons stabilizes 
antiferromagnetic order along this direction. 
Finally, it is also interesting to note that the optimized lattice constant $c$ for $n=1$ is shorter than that for $n=2$ 
(see Tab.~\ref{table_lattice}), which is also qualitatively in good agreement with experimental observations.~\cite{may@nm09}

\begin{table}[htbp]
\caption{\label{table_lattice} The optimized lattice constant $c$ (averaged value within the supercell and 
in unit of  \AA{}) and $c/a$ of (LaMnO$_{3}$)$_n$/(SrMnO$_3$)$_{2n}$ 
superlattices ($n=1,2$) calculated using GGA. The magnetic structures are A-type and C-type antiferromagnetic 
for SrTiO$_3$ and LaAlO$_3$ substrates, respectively. 
 }

\begin{tabular}{|c|c|c|c|c|}
\hline
  \textit{n} & \multicolumn{2}{|c|}{SrTiO$_3$ substrate}  & \multicolumn{2}{|c|}{LaAlO$_3$ substrate}\\\cline{2-5}
  &$c$&$c/a$&$c$&$c/a$\\  \hline
\textit{n}=1&   3.806   &  0.9746  & 4.006&1.0115\\
\hline
\textit{n}=2&    3.825  &  0.9795 & 4.010&1.0525\\  
\hline
\end{tabular}
\end{table}

\subsection{\label{results_lao}(LaMnO$_3$)$_n$/(SrMnO$_3$)$_{2n}$ on LaAlO$_3$}

Now, let us study the electronic and the magnetic properties of (LaMnO$_3$)$_n$/(SrMnO$_3$)$_{2n}$ superlattices ($n=1,2$) 
on (001) LaAlO$_3$ substrate. In the alloy manganites La$_{1-x}$Sr$_x$MnO$_3$, it is known that $c/a$ is a key 
parameter in determining the magnetic ground states.~\cite{terakura}  
Here, we demonstrate that even in these superlattices the 
magnetic structure can be controlled by the substrate strain which varies $c/a$.

\begin{figure}[ht]
\begin{center}
\includegraphics[width=20pc]{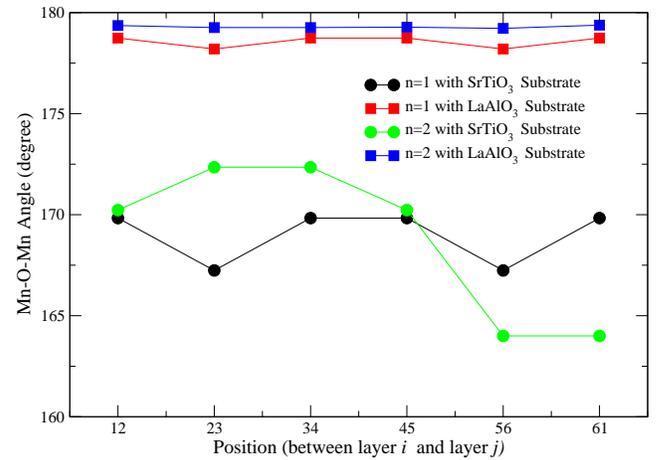}
\end{center}
\caption{ (Color online). 
Mn-O-Mn angles between the nearest layers along the $c$ direction for the relaxed crystal structures (calculated using GGA) 
for (LaMnO$_{3}$)$_n$/(SrMnO$_{3}$)$_{2n}$ (001) superlattices grown on different substrates indicated in the figure. 
The layer positions in the horizontal axis are indicated in Fig.~\ref{orbital} (a) and Fig.~\ref{orbital2} (a). 
}
\label{angle}
\end{figure}

\begin{figure}[ht]
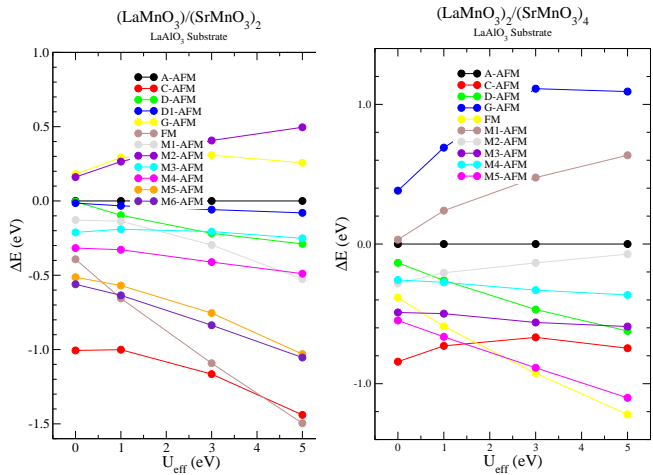

\begin{center}
\includegraphics[width=10pc]{La1Al-ground.eps}
\includegraphics[width=10pc]{La2Al-ground.eps}
\end{center}
\caption{\label{La-Al}  (Color online). 
$U_{\rm eff}$ dependence of the relative energies (calculated using GGA+$U$) for various magnetic structures 
(see Fig.~\ref{crystal-La1} and Fig.~\ref{crystal-La2}) compared to A-type antiferromagnetic state for 
(LaMnO$_{3}$)$_n$/(SrMnO$_{3}$)$_{2n}$ with $n=1$ (left) and 
$n=2$ (right) on LaAlO$_3$ substrate. 
}
\end{figure}

Because the in-plane lattice constant of LaAlO$_3$ is much smaller than that of LaMnO$_3$ (bulk lattice parameter is 
3.935$\AA{}$), 
it is expected that 
the LaAlO$_3$ substrate induces compressive strain. In fact, we find in Tab.~\ref{table_lattice} that the lattice constant $c$ in the 
superlattices is larger than the in-plain lattice constant $a$. As a result of this tetragonal distortion, Mn $e_g$ orbitals are split and $d_{3z^2-r^2}$ orbital 
is lower in energy than $d_{x^2-y^2}$ orbital, which thus induces $d_{3z^2-r^2}$ orbital order. A signature of this orbital order can be 
seen in the projected charge density distributions shown in Fig.~\ref{orbital} (c) and Fig.~\ref{orbital2} (c). 
Because of this orbital order, the magnetic ground state is expected to be C-type antiferromagnetic. 
Considering $10-12$ different candidates for possible magnetic structures as shown in Figs.~\ref{crystal-La1} and \ref{crystal-La2}, 
our GGA calculations find that the ground 
states for $n=1$ and 2 are both C-type antiferromagnetic metals [Fig.~\ref{crystal-La1} (b) and Fig.~\ref{crystal-La2} (b)]. 
This magnetic alignment can be indeed clearly seen in the projected spin density distribution as shown in Fig.~\ref{orbital} (e) and 
Fig.~\ref{orbital2} (e). It is also interesting to note that the lattice distortion along the $c$ direction is less pronounced for the case 
of LaAlO$_3$ substrate as compared to the case of SrTiO$_3$ substrate. As shown in Fig.~\ref{angle}, Mn-O-Mn angles between the 
nearest layers along the $c$ direction for the superlattices on LaAlO$_3$ substrate is almost $180^\circ$, which certainly favors the ferromagnetic 
double exchange along this direction. 
We also find that the C-type magnetic structure is robust against electron 
correlations in Mn $d$ orbitals up to $U_{\rm eff}=4$ eV for $n=1$ and $U_{\rm eff}=1.5$ eV for $n=2$ (see Fig.~\ref{La-Al}).~\cite{note}

As in the case of SrTiO$_3$ substrate, we can discuss 
the N\'{e}el temperature $T_N$ for the C-type antiferromagnetic order by calculating the total energy, and the results are 
summarized in Tab.~\ref{table_energy}. Simply by comparing the total energies of the C-type and the A-type antiferromagnetic 
states, the difference of which gives a rough estimate of an effective magnetic exchange $J_{\rm eff}$, we find that 
the stabilization energy of the C-type 
antiferromagnetic state, i.e., $J_{\rm eff}$, is larger for $n=1$ than for $n=2$. This implies that $T_N$ for $n=1$ is higher 
than that for $n=2$. 
Since (LaMnO$_3$)$_n$/(SrMnO$_3$)$_{2n}$ superlattices ($n=1,2$) on (001) LaAlO$_3$ substrate have not been studied 
experimentally, these results provide the theoretical prediction which should be tested experimentally in the future.

\subsection{\label{confine}Confinement potential}

\begin{figure}[ht]
\begin{center}
\includegraphics[width=20pc]{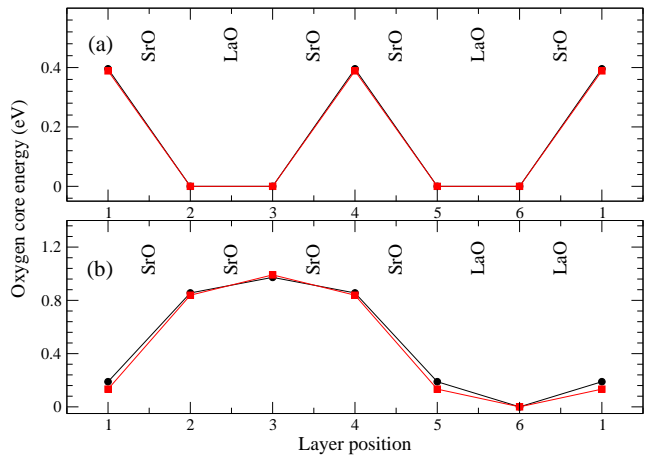}
\end{center}
\caption{(Color online). \label{core}The variations of the relative oxygen \textit{1s} core energy (calculated using GGA) 
in each MnO$_2$ layer of (a) 
LaMnO$_{3}$/(SrMnO$_{3}$)$_2$ and (b) (LaMnO$_{3}$)$_2$/(SrMnO$_{3}$)$_4$ superlattices. Results for SrTiO$_3$ 
and LaAlO$_3$ substrates are indicated by black circles and red squares, respectively. The layer positions are indicated 
in Fig.~\ref{orbital} (a) and Fig.~\ref{orbital2} (a). 
}
\label{potential}
\end{figure}

Finally, let us briefly discuss why the magnetic and orbital ground states found here are spatially uniform, in spite of 
apparent periodic potential modulation caused by different ionic charges, i.~e., 
La$^{3+}$ in LaMnO$_3$ layers, and Sr$^{2+}$ in SrMnO$_3$. 
As reported in Ref.~\onlinecite{nanda@prb09}, one way to estimate the effective potential modulation is to evaluate the 
oxygen 1$s$ core energy level. The results for (LaMnO$_3$)$_n$/(SrMnO$_3$)$_{2n}$ superlattices with $n=1$ and 2 
are shown in Fig.~\ref{potential}. From these figures, we see that (i) the potentials are almost the same for both substrates, and 
(ii) as is expected, the confinement potential becomes larger with $n$. The calculated charge density shows that $e_g$ 
electrons in LaMnO$_3$ layers is $\sim$0.2 (0.1) more than that in SrMnO$_3$ layers for $n=2$ ($n=1)$. This suggests that the 
thickness is still thin enough not to confine $e_g$ electrons in LaMnO$_3$ layers. 
However, we naturally expect that the bulk properties may recover far away from interface when $n$ is increased further 
and a metal-insulator transition should eventually occur. 

\begin{figure}[ht]
\begin{center}
\includegraphics[width=20pc]{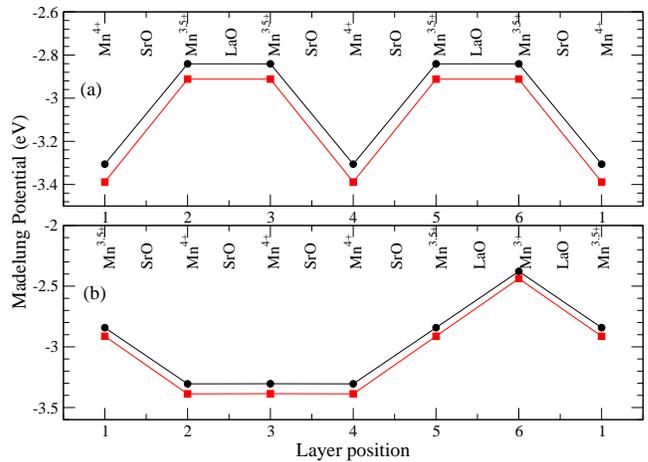}
\end{center}
\caption{(Color online). \label{core} 
The variation of Madelung potential for Mn ions in each MnO$_2$ layer of (a) LaMnO$_{3}$/(SrMnO$_{3}$)$_2$
and (b) (LaMnO$_{3}$)$_2$/(SrMnO$_{3}$)$_4$ superlattices. 
Results for SrTiO$_3$ and LaAlO$_3$ substrates are indicated by black circles and red squares, respectively.  
Here, the ideal crystal structures with no distortion, and the ideal Mn valency (indicated in the figures) with 
O$^{2-}$, La$^{3+}$, and Sr$^{2+}$ are assumed. The layer positions are indicated 
in Fig.~\ref{orbital} (a) and Fig.~\ref{orbital2} (a). Note that the sign convention of Madelung potential used here is 
that electrons prefer to locate in LaMnO$_{3}$ layers. 
}
\label{madelung}
\end{figure}

It is also found that the confinement potential can be more easily estimated simply by calculating Madelung potential. 
As shown in Fig.~\ref{madelung}, Madelung potential can indeed semi-qualitatively 
reproduce the values estimated from oxygen 1$s$ core energy level. This finding should be very useful in estimating  
the confinement potential for more complex superlattices in which first-principles electronic structure calculations are 
computationally expensive.

\section{\label{summary}Summary}

Using first-principles calculations based on the density functional theory, we have studied the effects of epitaxial strain on the magnetic ground states in 
(LaMnO$_3$)$_n$/(SrMnO$_3$)$_{2n}$ (001) superlattices with $n=1,2$. 
Our results clearly demonstrate that as in alloy manganites, even in superlattices, 
the epitaxial strain induced by substrates 
enforces tetragonal distortion, which in turn governs the ground state magnetic structure via the inherent orbital ordering. 
We have found that for the tensile strain induced by SrTiO$_3$ (001) substrate, the ground state is A-type antiferromagnetic 
metal with $d_{x^2-y^2}$ orbital order. The approximate estimation of an effective magnetic exchange suggests that the 
N\'{e}el temperature $T_N$ of the A-type antiferromagnetic order is higher for $n=1$ than that for $n=2$. 
These results are in excellent agreement with experimental observations.~\cite{may@nm09} 
Furthermore, we have predicted that 
for the compressive strain induced by LaAlO$_3$ (001) substrate, the ground state is C-type antiferromagnetic 
metal with $d_{3z^2-r^2}$ orbital order with higher N\'{e}el temperature $T_N$ for $n=1$ than that for $n=2$. 
These predictions should be confirmed experimentally in the future.

\section*{Acknowledgments}
The computation is done using the RIKEN Integrated Cluster of Clusters (RICC). Q.~Z. was supported by 
the Natural Science Foundation of Jiangsu Province (BK2012248), and research fund 
of Key Laboratory for Advanced Technology in Environmental Protection of Jiangsu Province (AE201152). S.~D. was supported by 
the 973 Projects of China (2011CB922101), NSFC (11004027), and NCET (10-0325). B.~W. was supported by NSFC (11174242).

{}


\begin{thebibliography}{}
\bibitem{note}
Note that with the GGA$+U$ scheme a Stoner type ferromagnetic state is expected stable 
for large values of $U_{\rm eff}$.
\end{thebibliography}
\end{document}